\documentclass[conference]{IEEEtran}
\usepackage{balance}      
\usepackage{url}
\usepackage{graphics}      
\usepackage[T1]{fontenc}  
\usepackage{mathptmx}
\usepackage{color}
\usepackage{booktabs}
\usepackage{textcomp}
\usepackage{enumitem}
\usepackage{cite}
\usepackage{dirtytalk}
\usepackage[super]{nth}
\usepackage{makecell}
\usepackage[dvipsnames]{xcolor}
\usepackage[most]{tcolorbox}
\usepackage[font=footnotesize]{caption}
\usepackage[utf8]{inputenc} 
\usepackage{microtype}    
\usepackage{ccicons} 
\usepackage{array}
\usepackage{multirow}
\usepackage{graphicx}
\usepackage{float}
\usepackage[framemethod=tikz]{mdframed}

\mdfdefinestyle{myFigureBoxStyle}{backgroundcolor=yellow!10,, roundcorner=25pt,tikzsetting={draw=black, line width=1pt}}%

\makeatletter
  \newcommand\fs@myRoundBox{\def\@fs@cfont{\bfseries}\let\@fs@capt\floatc@plain
  \def\@fs@pre{\begin{mdframed}[style=myFigureBoxStyle]}%
  \def\@fs@mid{\vspace{\abovecaptionskip}}%
  \def\@fs@post{\end{mdframed}}\let\@fs@iftopcapt\iffalse}
\makeatother

\usepackage{enumitem}

\IEEEoverridecommandlockouts

\def\BibTeX{{\rm B\kern-.05em{\sc i\kern-.025em b}\kern-.08em
    T\kern-.1667em\lower.7ex\hbox{E}\kern-.125emX}}
    
\makeatletter
\def\url@leostyle{%
  \@ifundefined{selectfont}{
    \def\UrlFont{\sf}
  }{
    \def\UrlFont{\small\bf\ttfamily}
  }}
\makeatother
\begin{document}

\title{Ask the Experts: What Should Be on an IoT Privacy and Security Label?}

\author{
\IEEEauthorblockN{Pardis Emami-Naeini}
\IEEEauthorblockA{Carnegie Mellon University \\
pardis@cmu.edu}
\and
\IEEEauthorblockN{Yuvraj Agarwal}
\IEEEauthorblockA{Carnegie Mellon University \\
yuvraj@cs.cmu.edu}
\and
\IEEEauthorblockN{Lorrie Faith Cranor}
\IEEEauthorblockA{Carnegie Mellon University \\
lorrie@cmu.edu}
\and
\IEEEauthorblockN{Hanan Hibshi}
\IEEEauthorblockA{Carnegie Mellon University \\
hibshi@andrew.cmu.edu}
}

\maketitle

\begin{abstract}
Information about the privacy and security of Internet of Things (IoT) devices is not readily available to consumers who want to consider it before making purchase decisions. While legislators have proposed adding succinct, consumer accessible, labels, they do not provide guidance on the content of these labels. In this paper, we report on the results of a series of interviews and surveys with privacy and security experts, as well as consumers, where we explore and test the design space of the content to include on an IoT privacy and security label. We conduct an expert elicitation study by following a three-round Delphi process with 22 privacy and security experts to identify the factors that experts believed are important for consumers when comparing the privacy and security of IoT devices to inform their purchase decisions. Based on how critical experts believed each factor is in conveying risk to consumers, we distributed these factors across two layers---a primary layer to display on the product package itself or prominently on a website, and a secondary layer available online through a web link or a QR code. We report on the experts' rationale and arguments used to support their choice of factors. Moreover, to study how consumers would perceive the privacy and security information specified by experts, we conducted a series of semi-structured interviews with 15 participants, who had purchased at least one IoT device (smart home device or wearable). Based on the results of our expert elicitation and consumer studies, we propose a prototype privacy and security label to help consumers make more informed IoT-related purchase decisions. 
\end{abstract}


\section{Introduction} \label{sec:introduction}
With the rapid growth in the development and deployment of IoT devices worldwide, large numbers of privacy and security issues have come to light~\cite{sivaraman2015network}. The popular press has reported on incidents such as Amazon Alexa sending private conversations to a random phone contact~\cite{alexa_out} and baby monitors getting hacked~\cite{foscam}. Even well-known IoT manufacturers have faced criticism for lack of transparency about their data practices. For instance, recent news articles reported that Google forgot to mention that its Nest Secure hub has a microphone~\cite{google} and Amazon revealed that in addition to artificial intelligence algorithms, human employees listen to a subset of audio recordings from Echo devices~\cite{alexa}.

Surveys have indicated that consumers are indeed worried about the collection of their personal data and who it is being sold to or shared with~\cite{conc1, conc2}. Privacy is of particular concern to users of IoT devices. Mozilla surveyed 190,000 participants around the world and found that privacy is the biggest IoT-related concern for 45\% of the respondents~\cite{moz}. In a survey conducted by the Economist Intelligence Unit (EIU) of over 1,600 IoT consumers, 92\% of participants were concerned about their privacy and wanted to have control over personal information collected by smart devices~\cite{EIU, phelps2000privacy}.    

While consumers are increasingly interested in purchasing smart devices~\cite{smarthome,marketIoT}, a recent survey conducted in the United Kingdom showed that security and privacy are key pieces of information that participants would like to consider when purchasing them~\cite{DCMS}. Similarly, an interview-based study of IoT consumers reported that almost all participants desired to have information about the privacy and security of IoT devices at the time of purchase~\cite{emami2019exploring}. Currently, there is little public information about the privacy and security of IoT devices that consumers can access to inform their purchase decisions~\cite{blythesecurity}. Some resources, such as the Mozilla \say{Privacy Not Included} website~\cite{moz_include} and a report published by the UK Information Commissioner's Office~\cite{christmas}, provide information about specific IoT devices or address limited privacy and security factors. 

Critical privacy and security information could be provided to consumers by including it prominently on a privacy and security label accompanying the device. This could also increase consumers' trust in the device manufacturer~\cite{walker2016surrendering}. In a May 2019 proposal, the UK Digital Ministers declared their intention to mandate security labels for IoT devices, with the goal of notifying consumers about security aspects of these devices~\cite{ukgov}. However, this plan only covers three security practices: using no default passwords, having a vulnerability disclosure program in place, and specifying the lifetime of security updates. Other proposals for IoT privacy and security labels fail to specify the specific information that consumers should be presented with on the label~\cite{FTC, NTIA, uk, congress, senate, ec, hub}. 

Given consumers' scarce attention, presenting them with the most relevant security and privacy information in the most digestible form is crucial. To determine the most important information to include on IoT privacy and security labels, we solicited the opinions of privacy and security experts. In various fields, expert elicitation has been used effectively for research and decision-making~\cite{veen2017proposal, knol2010use, usher2013expert, sheng2009improving, balebako2015notice}, particularly under uncertainty and when necessary information cannot be obtained from other sources~\cite{colson2018expert, kuhnert2010guide, hibshi2018composite}. 

We conducted interviews and surveys with 22 privacy and security experts. To get different perspectives, we recruited experts from industry, academia, government, and non-governmental organizations (NGOs). We also ensured that these experts come from different backgrounds related to IoT (software, hardware, and policy). We used the iterative Delphi methodology (explained further in Section~\ref{sec:methodology}) to develop a consensus among the experts around important factors and an understanding of their reasons for or against including each factor. Overall, we found that differences in opinions were driven less by fundamental differences in beliefs, but rather by differences in work experience and priorities. For example, some experts were more knowledgeable about specific security mechanisms, standards, or regulations, and prioritized factors related to their area of expertise or their organization's mission. Prior research has shown that security experts might analyze the same artifacts differently depending on their background in specific security domains~\cite{Hibshi2016agrounded}. 

Most factors identified as important by experts are factors that they believe will inform consumers. Experts also identified some factors for inclusion that could inform experts only, mostly to be able to hold companies accountable.

Prior studies suggest that layered labels can be effective~\cite{emami2019exploring, layer1, schaub2015design, cranor2012necessary}. A layered label includes a \textit{primary} layer that presents the most important and glanceable content, followed by a \textit{secondary} layer for additional information. In our study, we asked experts to specify the layer on which the information should be included on the label. They mostly recommend putting only information that would be understandable and important to most consumers on the primary layer.  

We designed a prototype layered label based on our expert elicitation study. We then conducted semi-structured interviews with 15 consumers of IoT devices (smart home devices or wearables) and presented our prototype to them. We show that all of our participants had a clear understanding of the information presented on the primary layer of the label. Although some of the factors on the secondary layer of the label were less understandable to participants who lacked privacy and security expertise, all of our participants reported that they still want such information to be included on the label mainly to be as informed as possible. In addition, consumer participants reported that having all the important privacy and security factors, even unfamiliar ones, on the label would help them easily search online to find more information. 

We make the following contributions in this paper: 
\begin{itemize}
\item We distill an extensive list of privacy and security factors to identify the most important pieces of information to include on IoT labels.
\item We partition the most important factors into two layers:  the primary factors we want consumers to notice and consume at a glance, and the secondary factors that require more space to effectively convey risk to consumers.
\item Based on our expert and consumer interviews and surveys, we propose a prototype IoT label that includes the most important factors with proposed groupings.
\end{itemize}

\section{Background and Related Work} 
In this section, we first outline prior research on understanding consumers' privacy and security related concerns for IoT devices. Next, we provide background on how labels have been used in different contexts to inform consumers' choices. Finally, we discuss reports and recommendations on privacy and security best practices for consumer IoT devices.  

\subsection{Consumers' Privacy and Security Concerns and Preferences}
Numerous studies have demonstrated that consumers are increasingly concerned about the privacy and security of IoT devices~\cite{granjal2015security, arias2015privacy, sicari2015security, moz}. Researchers have shown that the extent of these concerns depends on factors such as the type of data collected, the purpose of data collection, and the retention of collected data. Lee and Kobsa~\cite{lee2016understanding, lee2017privacy} explored factors related to IoT data collection and found that people are more concerned about data being collected in private locations compared to public locations. In addition, they found that people's concerns depend on the type of data being collected. In a previous study Naeini et al. used vignettes to study how different factors can explain variations in comfort level related to data collected by sensors. They found that factors such as the purpose of data collection and the retention time significantly impact people's privacy-related concerns~\cite{naeini2017privacy}. 

Despite such concerns, consumers are still purchasing IoT devices, mostly due to their perceived convenience and features~\cite{cisco}. This is sometimes referred to as a \say{privacy paradox,} thanks to the discrepancy between privacy concerns and actions taken to mitigate those concerns~\cite{acquisti2015privacy}. One cause for this could be that consumers are provided with little, or often no, privacy and security information about IoT devices prior to purchase~\cite{uk, emami2019exploring, blythesecurity, blythe2018consumer}. This prevents consumers from making informed IoT-related purchase decisions and increases the risk of privacy and security vulnerabilities, which may lead to high-profile and large-scale attacks targeting IoT devices~\cite{antonakakis2017understanding}.

\subsection{Product Labels}
Product labels, such as food nutrition and energy labels, have been used to aid consumers' purchase decisions. Food nutrition labels, in particular, were developed to decrease obesity by helping consumers purchase healthier food products~\cite{draper2011front}. Other objectives of food nutrition labels include encouraging food companies to compete to produce healthier products and allowing governments to support consumers' health-related behaviors without mandating specific nutritional requirements~\cite{kleef2015growing}. 

The effectiveness of food nutrition labels has been shown to depend on factors such as consumers' attention at the point of sale~\cite{szanyi2010brain}, whether the consumer is purchasing the product for children, aiming to lose weight, and/or purchasing the product for the first time~\cite{schor2010nutrition, vyth2010front}, their nutrition-related knowledge, health condition~\cite{crites2005impact, drichoutis2005nutrition}, and socio-demographic factors. Studies have also shown the limitations of nutrition labels in effectively communicating nutrition information to consumers to improve eating habits~\cite{nut1, nut2}. Despite these shortcomings, these studies demonstrate that food nutrition labels significantly inform consumers' purchase decisions~\cite{miller2012making, nayga2000nutrition}. 

In the realm of privacy, researchers have explored the impact of privacy ``nutrition labels'' on websites. They found that privacy labels help users find important information faster and more accurately, as compared to finding such information in traditional privacy policies~\cite{kelley2010standardizing}.   

\subsection{Privacy and Security Guidelines and Best Practices}
Tanczer et al.~\cite{tanczer2018summary} conducted an extensive literature review on publicly available reports from industry associations and international organizations on their security-related proposals and recommendations for consumer IoT devices. They reviewed 17 industry reports (including from Intel, HP, and Consumer Technology Association) and policy reports (including from European Commission, International Organization for Standardization (ISO), and Alliance for the Internet of Things Innovation (AIOTI)). This review observed 19 overarching principles related to security best practices that were referred to at least twice in these reports. The most common principles (mentioned in at least 10 reports) were strong authentication by default, reliable and cryptographically signed security updates, encryption by default, and compliance and risk assessment. Some of the other principles were related to physical security, vulnerability reporting and disclosures, and secure device boot. The security factors that we synthesized based on our expert elicitation study covered all of the most frequent security principles mentioned in this literature review~\cite{tanczer2018summary}.

Tanczer et al. concluded that in general, the industry acknowledges the importance of selling safe and secure IoT devices in the market and would like to work alongside the government as part of their efforts. However, they are more interested in self-regulation than in government interventions~\cite{tanczer2018summary}. For example companies can self-certify their IoT devices using a framework developed by IoT Security Foundation (IoTSF) that specifies five levels of compliance~\cite{IoTSF}.

A recent UK government report argued against self-regulation, noting the lack of incentives for IoT companies to adhere to security best practices when designing IoT products. The report recommended that the government mandate specific requirements for IoT devices as a mechanism to improve the security of consumer IoT products~\cite{mandate}. These requirements are no default password, availability of a vulnerability disclosure program, and security updates. These recommended requirements all are included in our privacy and security label (see Figures~\ref{fig:primary} and~\ref{fig:secondary}).

Notably, all the reviewed reports above focused on devices' security mechanisms with few references to data privacy considerations. As consumers are concerned about both the privacy and security of their devices, in our work we asked experts and consumers about both privacy policies and security mechanisms that should be included on a label.

\section{Methodology} \label{sec:methodology}
We first conducted an expert elicitation study to specify the content of a privacy and security label for IoT devices. We complemented the expert study with a series of 15 semi-structured interviews with non-expert consumers and iterated on the label design.

\subsection{Expert Elicitation Study}
\begin{figure*}[t]
\includegraphics[trim={2.5in 10.3in 9.35in 2.6in},clip,width=\textwidth]{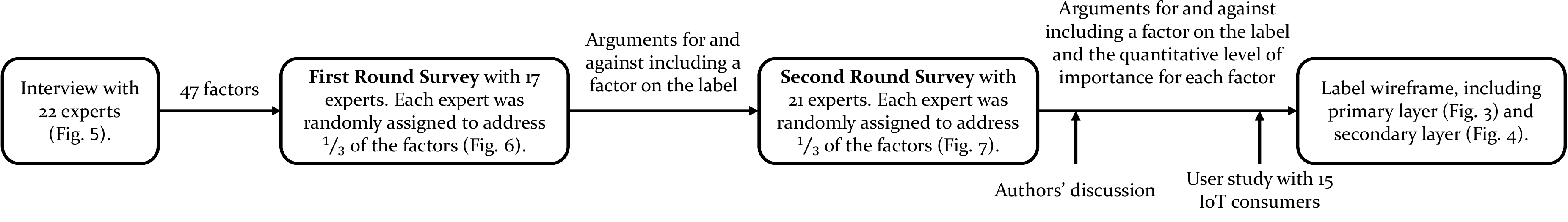}
    \caption{We followed a three-round Delphi method, by conducting an interview study and two rounds of surveys. Finally, we designed a label prototype that captures the findings of the process, inputs from authors' multiple rounds of discussion, and a user study with 15 IoT consumers.}
    \label{fig:flow}
\end{figure*}

In the expert elicitation study, our overarching goals were to identify factors that experts believed would be useful to include on a privacy and security label for IoT devices and to understand the experts' rationale for selecting each factor. We conducted an in-depth, semi-structured interview study, followed by two rounds of surveys with 22 privacy and security experts. The process is depicted in Figure~\ref{fig:flow}.

\subsubsection{Participant Recruitment and Compensation} \label{sec:exp_rec}
To capture a wide range of expert opinions, we recruited  experts from academia, industry, government, and non-governmental organizations (NGOs) in the United States, with a diverse range of expertise: software, hardware, policy, standards, and user experience (UX). We recruited experts with whom the authors had interacted professionally or were recommended by other experts. We carefully selected experts, who are all well-known in their respective fields. More specifically, we looked for experts, who satisfied at least one of the following qualification criteria. Seven experts met two criteria. 

\begin{itemize}
    \item Computer science or engineering professor in the field of privacy and security.
    \item More than 10 years of research or practice in the field of privacy, security, or policy.
    \item Author of notable books in the field of privacy and security.
    \item Active involvement in cybersecurity standardization.
    \item Leading a corporate IoT product team.
\end{itemize}

After identifying the experts who met the qualifications we were looking for, we contacted them and invited them to either come to our institution for an in-person interview or join our interview study online over Skype. All the interviews were audio recorded and transcribed by a third-party service. We compensated experts with a \$25 Amazon gift card.

\subsubsection{Delphi Method}
\label{sec:delphi}
As defined by Delbecq et al., the \textit{Delphi method} is \say{a method for the systematic solicitation and collection of judgments on a particular topic through a set of carefully designed sequential questionnaires interspersed with summarized information and feedback of opinions derived from earlier responses}~\cite{atherton1976group}. This method of qualitative research was originally developed by Dalkey and Helmer in the 1950s and has been widely used to reach consensus between a group of experts without face-to-face interactions~\cite{dalkey1963experimental}. The Delphi method has been used in a number of studies related to policy design and implementation~\cite{adler1996gazing}, social science~\cite{strauss1975delphi}, and human-computer interaction~\cite{miaskiewicz2006use}.

The Delphi method has three important features. First, the responses as well as group interactions in each round are anonymized. Second, the process involves multiple rounds of data collection procedures (e.g., interview, survey), and finally, in each round, the summary of the previous round is shown to experts as a means to reach consensus~\cite{cochran1983delphi, cyphert1971delphi, dailey1990delphi}. The study continues until consensus is reached, which generally occurs after three iterations~\cite{ludwig1997predicting}.

\subsubsection{Expert Interviews}
The first phase of the Delphi method is open ended~\cite{murry1995delphi}. Therefore, our first step was to conduct semi-structured interviews with privacy and security experts. 

We began the interviews by introducing the idea of a privacy and security label and its similarity to a food nutrition label. Following the introduction to the study and its goals, we asked experts to provide their definition of privacy and security as it relates to IoT devices. Next, we asked experts to think about the content of the label and specify the information that they think should be on a privacy and security label for IoT devices. For each piece of information they specified, we asked them to consider whether it was relevant to consumers or experts.   
In an iterative process, we compiled a list of security and privacy factors suggested by the experts we interviewed, and added new factors suggested during each interview. Towards the end of each interview, we presented the full list of factors so that each expert reviewed their own factors, as well as the factors suggested by previously-interviewed experts (see Figure~\ref{fig:interview}).  

\subsubsection{First Round Survey}
The expert interviews resulted in an extensive list of privacy, security, and general factors that experts wanted to see on an IoT label. We then conducted a survey of the same experts to understand the rationale behind their preferences (see Figure~\ref{fig:survey1}). In order to decrease fatigue, we split the factors in the survey so that each expert was presented with one-third of all the factors. For each expert, the ordering of factors was randomized. 

When introducing the survey to experts, we explained that in a layered IoT label, the first or primary layer would include the most important information, and the secondary layer would contain the information on the primary layer as well as additional helpful information. We also advised experts not to worry about the design of the label when answering the questions. Then, for each factor, we asked the experts to specify whether they believe that factor is important to include on the label, and to provide reason(s) that support their answer. 

\subsubsection{Second Round Survey}
For agreement over the inclusion and exclusion of factors, we conducted a second survey with the same set of experts (see Figure~\ref{fig:survey2}). When introducing the second survey to the experts, we described our two key objectives for the content of the IoT label: to inform consumers, and to provide a means for holding companies accountable for their privacy and security practices. To reduce respondent fatigue, participants answered questions for one-third of all the factors, randomly chosen from the three categories---security, privacy and general---such that they saw approximately the same number of questions within each category. 

In this second survey, we used data collected from our interviews and first survey. We presented each factor from our dataset alongside the experts' reasons for inclusion or exclusion. Then, on a five-point Likert scale, we asked each expert to decide whether they believe the factor should be included on the label and to provide their rationale if different from what we presented. Next, we asked them to specify on which layer of the label they would like to place this factor and the rationale behind their choice. We asked them to classify the factor as being most relevant to privacy, security, or general information. Finally, we asked experts to provide any additional comments about the factor that came to mind. 

In addition to the questions we asked for each factor, we asked experts about their opinion on separating or merging privacy and security sections on the label. At the end of the survey, we asked experts to state their privacy and security expertise and domain of knowledge, followed by some general demographic questions.  

\subsubsection{Data Analysis}
We collected approximately 22 hours of interview audio recordings. We used thematic analysis to qualitatively summarize interview transcripts, following the approach suggested by Braun and Clarke~\cite{braun2006using}: 
\begin{itemize}
    \item Phase 1: A primary coder read the interview transcripts and took notes, listening to parts of the audio files as needed when the transcripts were incomplete.  
    \item Phase 2: The primary coder created the initial codebook by examining the notes from the interview phase and the notes from Phase 1 above, listening for reasons for and against including factors on an IoT label. This step did not focus on finding patterns in the responses. 
    \item Phase 3: The primary coder merged the smaller codes into broader themes. This step focused on finding patterns and themes from the long list of codes from Phase 2.   
    \item Phase 4: The themes that emerged from Phase 3 were reviewed and discussed by the authors of the paper to resolve any disagreements. This step helped increase the validity of the themes. In an iterative process, some of the themes were removed from the codebook and some themes were merged into more general themes until we achieved consensus on the final themes.   
    \item Phase 5: The finalized themes (reasons to include or exclude each factor) were moved into the final codebook.
\end{itemize}

The finalized privacy, security, and general factors were used as input to the first round of our survey, where we asked experts to provide us with their arguments. We then followed the same coding process described above to code the open-ended survey responses. After the first survey, we revised the themes (reasons for and against including a factor on the label) in the codebook and presented them to the experts in the second survey. 

We reached a point of saturation in terms of finding new factors after interviewing 20 experts. In other words, no new privacy, security, or general factors were mentioned by our participants in the rest of the interviews as well as the two follow-up surveys. 

Thematic analysis is purely qualitative and inductive ~\cite{braun2006using}. The literature showed that having more than one coder does not make the codes objective, since two coders could apply the same subjective perspective to the data~\cite{marks2004research}. Indeed according to a survey of CSCW and HCI publications from 2016 to 2018, only 6\% of papers using thematic analysis used multiple coders and measured Inter-rater reliability~\cite{mcdonald2019reliability}. 

An iterative, yet inductive, analysis approach in thematic analysis increases the reliability of the theme-finding process \cite{franklin2001reliability}. All the themes were iteratively and extensively discussed among the researchers in the group. For any disagreement, researchers traced the theme back to its corresponding subcodes and checked whether the source of disagreement was the subcodes that were used. If not, we traced the subcodes further back to experts' quotes from the transcriptions and we then decided on the appropriate subcodes and the appropriate themes arising from the subcodes. This iterative approach is recommended with qualitative methods that are high in subjectivity~\cite{franklin2001reliability, madill2000objectivity}. We also improved the reliability by consulting with the expert participants using the second round survey mentioned above; this triangulation method is known as testimonial validity or member checking~\cite{saldana2015coding}.

\subsection{Semi-Structured Interviews with Non-Expert Consumers}
We used the results of our expert study to inform the development of prototype designs for primary and secondary labels. We created prototype boxes for two fictitious brands of security cameras and included a primary-layer label on each box. We put the corresponding secondary-layer labels on a mock-up of an online shopping website. Next we conducted a semi-structured interview study with 15 non-expert consumers to gain insights into how they would use these labels and how well the labels convey risk.

\begin{figure}[t]
\centering
\includegraphics[angle=-90,trim={14.5in 10.3in 12in 3.6in},clip,width=.4875\textwidth]{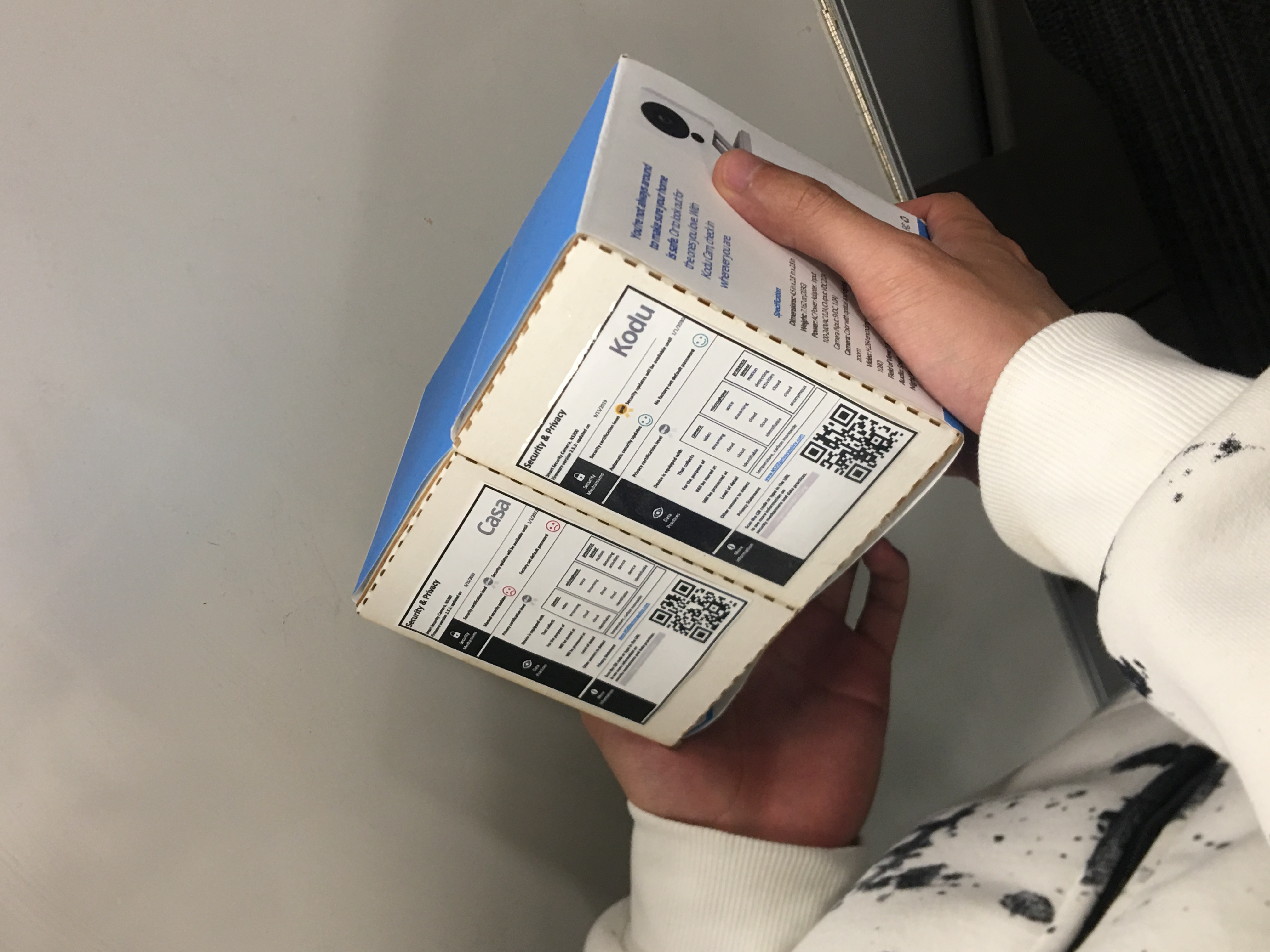}
    \caption{A user study participant comparing the privacy and security practices of two hypothetical smart security cameras.}
    \label{fig:compare_shop}
    \vspace{-1em}
\end{figure}

\subsubsection{Participant Recruitment and Compensation}
We recruited participants by posting on Craigslist, Reddit, and our institution's recruitment website. Participants were required to be at least 18 years old and have purchased at least one smart home device or smart personal device. Prospective participants completed a short screening survey, in which we collected demographics and asked about what IoT devices they had purchased and how they purchased them. We invited a diverse sample of qualified participants to our lab for a 1-hour interview. Each interviewee received a \$25 Amazon gift card.

\subsubsection{Initial Questions}
We showed participants a box for a hypothetical security camera that did not include a label and asked them what they could tell about the privacy and security of the device by looking at its box. We then asked participants whether they had ever seen an informative label on any product. Next, we presented them with one of the two labeled security camera boxes and asked them what they could tell about the privacy and security of the device. We asked a number of questions to study participants' understanding of the content of the label and whether the information conveyed risk.

\subsubsection{Risk Communication in Comparative Purchase Process}
We showed participants the other labelled security camera box, and told them this camera had the same price and features, but with different privacy and security information. We asked them to compare these two products and discuss which has better privacy and security, which device would they purchase, and the information that helped them make this decision (see Figure~\ref{fig:compare_shop}). 

We then told participants about the secondary layer of the label, which can be accessed by scanning the QR code or typing in the URL on the first layer. After introducing the idea of the layered label, we asked participants whether they had ever seen one on any other product. We asked them to discuss the pros and cons of a one-layer and two-layer label. 

\subsubsection{Information Comprehension in Non-comparative Purchase Process}
We asked  participants to look at the information on the label of the product they decided not to purchase  and to discuss their concerns and their understanding of the information.

\subsubsection{Risk Communication in Non-comparative Purchase Process}
We asked participants to specify the factors that seemed risky to them from a privacy and security perspective and discuss what kinds of risk they would be exposed to. We also asked how the product could be improved to reduce this risk.

\subsubsection{Secondary-Layer Information Comprehension}
We asked participants whether they would prefer to scan the QR code or type in the URL to look for additional information. Based on their preference, we scanned the QR code or typed in the URL on the primary layer to show the secondary layer to participants. We then asked them to start from the beginning of the label and tell us what each factor means to them, how useful they believe each factor would be, and if they have any suggestions to make the information more understandable.  

\subsubsection{Label Format}
We asked questions about the label format, including the separation of factors into privacy, security, and general information sections. We also asked participants to specify the factors that they believed are currently misplaced, and should be either removed from the label or moved to another section or layer of the label.

\subsubsection{Purchase Behavior}
Finally, we asked questions to understated participants' purchase behavior related to online and in store shopping.

\subsubsection{Data Analysis}
We collected about 15 hours of audio recordings, which we had transcribed. The first author was the primary coder who created the codebook and kept it updated throughout the coding process. To analyze the data, we used structural coding, which is  appropriate for coding semi-structured interviews~\cite{saldana2015coding, kathleen2008team}. We defined four structural codes (e.g., attitudes toward layered labels, reasons to include or exclude a factor from the label), which we divided into 13 subcodes (e.g., being as informed as possible, lack of relevance to privacy and security). Unlike thematic analysis, structural coding is more objective, and results in a codebook used for categorization~\cite{saldana2015coding}. Therefore, having more than one coder and using inter-rater reliability is helpful in testing the reliability of the codebook~\cite{franklin2001reliability}. Each interview was independently coded by two researchers, who then discussed and iteratively revised the codebook. After resolving the coding disagreements, we reached the Cohen's Kappa inter-coder agreement of 84\%. Cohen's Kappa inter-coder agreement of over 75\% is considered as \say{excellent} rate of agreement~\cite{fleiss2013statistical}. In case of disagreement, we report on the results of the primary coder.

\subsection{Ethics}
Both expert and consumer studies were reviewed and approved by our Institutional Review Board (IRB). All participants provided their informed consent to participate in the surveys and interviews, to have audio recorded, and to have the recordings transcribed by a third-party transcription service.

\subsection{Limitations} 
Expert elicitation is prone to overconfidence and cognitive biases~\cite{speirs2010reducing}. To reduce overconfidence, in the second expert survey, we presented strong arguments for and against having each factor on the label so that experts could read the rationale provided by other experts before indicating their own preferences.

The experts interviewed in this study are not representative of the entire population of privacy and security experts. Our aim was to surface a wide variety of expert viewpoints. Therefore, we recruited experts with diverse expertise related to IoT security and privacy and from different sectors. We selected experts based on our inclusion criteria, as discussed in Section~\ref{sec:exp_rec}.

In the follow-up surveys with experts, we presented each participant with only one-third of factors randomly sampled from the list of all privacy and security factors. This reduced respondent fatigue~\cite{backor2007estimating} and increased the quality of responses, at the cost of not being able to achieve a true consensus across all experts. In the expert study, our main objective was to collect the opinions of diverse experts, and not to reach perfect consensus. Therefore, we report themes that were only mentioned by a few experts. 

We designed a label prototype based on findings from our three-round Delphi process. However, we did not conduct a fourth round of study to show experts the label and ask them for feedback. Although this would have helped us confirm experts' opinions about the factors in the context of a complete label design, it would have introduced confounding factors related to the design of the label, including, but not limited to, the order of sections on the label and the specific language used to convey the information. Since the expertise of the participants in our study was in the area of IoT security and privacy, and not in communications design, we limited the expert elicitation study to focus on the individual factors.

Our consumer study was a small-scale qualitative study designed to gain initial consumer feedback and assess the overall usefulness of the layered label approach in this context. Additional large-scale iterative design and testing is needed to refine and validate the label design.

\section{Results} \label{sec:results}

\begin{table}[t]
\caption{We conducted an expert elicitation study with 22 privacy and security experts. NGO stands for non-governmental organization and UX stands for user experience.}
\centering\def\arraystretch{}
\footnotesize{
\setlength\tabcolsep{2.1pt}
\begin{tabular}{|c|c|c|c|}
\hline
\begin{tabular}[c]{@{}c@{}}Expert \\ ID\end{tabular} & \begin{tabular}[c]{@{}c@{}}Privacy \& Security \\ Expertise\end{tabular} & IoT Focus & Workplace \\ \hline
P1 & Privacy & Policy, standards & Enterprise \\
P2 & Privacy & Policy, UX & Enterprise, NGO \\
P3 & Privacy & Software & University \\
P4 & Privacy & Policy & University \\
P5 & Privacy & Hardware & Enterprise \\
P6 & Privacy & Policy, software, UX & Enterprise \\
P7 & Privacy & Policy & NGO \\
P8 & Privacy & Policy, privacy & NGO \\
P9 & Privacy & Policy, privacy & NGO \\
S1 & Security & Software & University \\
S2 & Security & Software & University \\
S3 & Security & Policy, software & Government, University \\
S4 & Security & Policy, security & Enterprise \\
S5 & Security & Hardware, security & Enterprise \\
S6 & Security & Software & Enterprise \\
S7 & Security & Policy & Enterprise, NGO \\
S8 & Security & Policy & NGO \\
S9 & Security & Hardware, software & Enterprise \\
B1 & Both & Software & University \\
B2 & Both & Policy, software & Enterprise \\
B3 & Both & Policy, standards & NGO \\
B4 & Both & Policy, software & Enterprise \\ \hline
\end{tabular}}

\label{tab:expert_demographic}
\vspace{-1em}
\end{table}

We conducted 22 one-hour, semi-structured interviews with IoT privacy and security experts with diverse backgrounds as described in Table~\ref{tab:expert_demographic}. We compiled a list of 47 privacy, security, and general factors that experts said they would like to see on the IoT label. 

We followed the expert interviews with a qualitative survey to understand the reasons experts wanted to include or exclude each factor. Out of 22 invited experts, 17 answered the first survey, with each of them being asked to comment on one-third of the 47 factors. This survey took an average of 16 minutes to complete. We collected on average seven reasons for or against including each factor on the label. We then conducted thematic analysis on the arguments provided to arrive at two or three primary reasons for and against each factor. 

In the second survey, we presented experts with the reasons for and against each factor from the previous two phases and asked them to rate their enthusiasm for including the factor on either the primary or the secondary layer of the label. 21 experts participated in the second survey, spending an average of 15 minutes. We identified 12 factors that most experts recommended including on the primary layer and 13 factors most experts recommended including on the secondary layer.

Based on the expert study and authors' discussions, we designed prototype privacy and security labels for hypothetical smart security cameras and presented them to a diverse sample of 15 non-expert consumers (see Table~\ref{tab:consumer_demographic}). We asked them questions related to their understanding of the factors on the label and whether they conveyed risk. We iteratively improved the content of the label to make it more understandable, resulting in a final prototype label.

In this section, we first discuss experts' attitudes toward privacy and security. Next we present the factors experts wanted to include on the primary and secondary layers of the label or exclude from the label, followed by a discussion of how consumers perceived those factors. We continue by discussing  consumers' attitudes toward the labels and their layered design. Finally, we present our prototype label design that we designed based on experts' and consumers' input. 

\newcolumntype{P}[1]{>{\centering\arraybackslash}p{#1}}
\begin{table}[t!]
\caption{User study participants, demographics, and devices they have purchased.}
\setlength{\tabcolsep}{0.3em} 
\centering\def\arraystretch{1}
\footnotesize{
\begin{tabular}{|l|l|l|c|m{5.7cm}|}
\hline
\multicolumn{1}{|c|}{\rotatebox[origin=l]{90}{Participant ID}} & \multicolumn{1}{c|}{\rotatebox[origin=l]{90}{Gender}} &
\multicolumn{1}{c|}{\rotatebox[origin=l]{90}{Age}} & \multicolumn{1}{c|}{\rotatebox[origin=l]{90}{\parbox{1.8cm}{Technical \\ Background}}} &
\multicolumn{1}{c|}{\multirow[t]{1}{*}[.31in]{IoT Devices our Participants Have Purchased}}  \\ \hline
C1 & F & 35-44 & Y & Thermostat, TV, switches, lock, outlet \\
C2 & F & 35-44 & Y & Vacuum cleaner, gaming consoles \\
C3 & M & 55+ & Y & Thermostat \\
C4 & F & 45-55 & Y & Smart speaker \\
C5 & M & 35-44 & Y & Camera \\
C6 & M & 55+ & N & Smart speaker, lights \\
C7 & M & 18-24 & N & Smart speaker, lights, TV \\
C8 & M & 25-34 & N & Activity tracker \\
C9 & F & 25-34 & N & Smart speaker \\
C10 & F & 25-34 & Y & TV, camera \\
C11 & F & 18-24 & N & Smartwatch, activity tracker, camera \\
C12 & F & 35-44 & N & Smart speaker \\
C13 & M & 25-34 & N & Smart speaker, TV, plugs, vacuum cleaner\\
C14 & F & 25-34 & N & Smartwatch, activity tracker, vacuum cleaner\\
C15 & F & 25-34 & Y & Smart speaker, smartwatch \\ \hline
\end{tabular}
}
\label{tab:consumer_demographic}
\vspace{-1em}
\end{table}

\subsection{Definition, Assessment, and Accountability}
We started the interviews by asking experts to define privacy and security related to IoT devices. When defining security, almost all experts (21/22) mentioned the CIA triad of confidentiality, integrity, and availability. However, experts had different definitions for privacy. Some experts (9/22) defined privacy as having transparency and control over data practices and some experts defined privacy as the confidentiality aspect of security (8/22). Overall, experts' definitions for security were mostly passive and focused on hardware and software enforcement mechanisms. On the other hand, their definitions for privacy were active and centered around policy, control, and individuals' preferences and comfort. For instance, P7 compared privacy and security practices by saying: \say{If the privacy is done right, it would be more active than security because the consumer would be able to be in control.} B3 explained how privacy and security are related by saying: \say{Security mechanisms are the things that enforce the technical controls that allow us the privacy we have.} 

Most experts (15/22) believed that security information is less tangible and understandable for consumers compared to privacy information, in part because it relies on technical mechanisms. S5 explained: \say{Consumers don’t necessarily understand some of the abstract stuff about security that they don’t see. Whereas when their privacy is breached, they are more aware of that.} 

In addition, almost all experts (19/22) reported that security practices are easier to measure and assess than privacy practices, as security is more objective and less controversial, while privacy is more subjective and context dependent. P5 explained that security is easier to quantify: \say{Security strikes me as less subjective, and, therefore, easier to measure. Which is to say that there could be certain standards. What sort of encryption exists on the device? What encryption is in the cloud? These are all fairly quantifiable. Whereas privacy is trickier, I think. And almost ethically and morally from my point of view, there's a lot more gray area in this.} This finding is aligned with the current efforts in IoT assessments and scoring, which are more focused toward security mechanisms than privacy practices~\cite{digital, score, GSMA, IoTSF, ENISA}. 

Experts reported that IoT privacy and security labels could increase accountability (most frequently mentioned for factors on the secondary layer) and transparency. Seven experts suggested that increased transparency could be an incentive for companies to compete on privacy and security, leading to safer products. S4 explained: \say{There is value in forcing the company to write a list down even if the consumer doesn't understand it. If you said, `list your open ports,' there would be an incentive to make them few.} 

Some experts (8/22) mentioned that IoT companies' accountability should be different for privacy and security breaches. They said that security breaches can happen accidentally, even if companies follow best practices. However, privacy violations could be intentional and IoT companies could even profit from them. P4 explained: \say{You can have the best intentions in the world, but if somebody comes up with some crazy hack overnight, you shouldn't be held responsible for it. You should be held responsible for fixing it. As opposed to if you intentionally share someone's data with a third party, it's not like, oh you could have prevented it but you chose to do it.} 

In the second survey, we asked experts to specify whether they prefer to see privacy and security factors in two separate sections or if we should combine them into one section. About half of the experts (10/21) believed privacy and security should be presented in separate sections. Most of these (9/10) said such separation would improve the readability and utility of the label and help educate consumers. P1 explained: \say{I lean toward the option of separation, because I'd like to see a streamlined label for most consumers to `consume' as easily and quickly as possible.} S8 concurred: \say{Consumers may have preferences for one aspect more than the other and stating them separately better enables consumer choice and education.}  However, the other half of the experts believed privacy and security factors should be combined into one section (11/21). For example, P4 believed that for some consumers, security seems more important than privacy. Thus, separating them on the label may cause consumers to focus only on security factors and ignore privacy information. Among those experts, who were more interested in combining privacy and security information into one section, almost all of them (9/11) mentioned that privacy and security are so related that it is not possible to completely separate these two concepts.

In the label that we presented to consumer interview participants, we grouped information into three main sections: security mechanisms, data practices, and general/more information. Participants preferred the proposed separation of sections and reported that these groupings made sense to them. 

\subsection{Factors to Include in the IoT Label} 
From the second survey, we found 30 factors that at least 4 out of 7 of the experts recommended including on the label (either on the primary or secondary layer)  and 17 factors that at least 4 out of 7 experts recommended excluding from the label. Note that since only a third of the factors were shown to each expert (total 21 on the second survey), at least 4 responses constituted a majority.  

The authors discussed the experts' arguments and preferences for each factor and made a decision as to whether or not each factor should be included and if so, on which layer of the label. In some cases, we made a decision that contradicted the majority of experts if we felt that their arguments could be accommodated in a different way. 

\subsubsection{Primary Layer}
We found 12 factors that at least 4 out of 7 experts wanted to include on the primary layer:

\begin{itemize}[leftmargin=*]
    \item {Privacy rating for the device from an independent privacy assessment organization}
    \item {Security rating for the device from an independent security assessment organization}
    \item {The date until which security updates will be provided}
    \item {Type of data that is being collected}
    \item {Type of sensor(s) on the device}
    \item {Whether or not the device is getting cryptographically signed and critical automatic security updates}
    \item {Types of physical actuations (e.g., talking, blinking) the device has and in what circumstances they are activated}
    \item {Whether or not the device is using any default password}
    \item {Frequency of data sharing (e.g., continuous, on demand)}
    \item {The warranty period of the device}
    \item {Level of detail (granularity) of the data being collected, used, and shared (e.g., identifiable, aggregate)}
    \item {Access control for device and apps (e.g., none, single-user account, multi-user account)}
\end{itemize}
 
Experts were interested in including these factors on the primary layer because they considered them necessary for consumers to know, they convey critical information about the privacy and security of the device, and they inform consumers' purchase decisions. For example, P1 explained why the type of collected data should be included on the primary layer: \say{I think this is the most useful information to be provided to consumers for them to compare privacy risks of IoT devices.}  

All of our consumer participants understood the information presented on the primary layer and were able to talk about privacy and security implications of each factor. For example, consumers associated the expiration date of the device to its security updates lifetime. C3 mentioned \say{planned obsolescence} when talking about the security update lifetime: \say{I do like the fact that you say when the security updates will no longer be available, because that alerts people to the fact that this device is going to expire. People have thermostats that last for decades and it's useful to know that this is planned obsolescence.} One of our participants brought up a point of skepticism, related to how long a company claims to provide security updates: \say{I'm skeptical because I know that tech startups can very rarely guarantee that their servers will be online for three or more years.}

Among the factors experts believed should be included on the primary layer, there were three factors that we decided to either move to the secondary layer or exclude from the label. Note that we also removed privacy and security star ratings later in the process as mentioned in Section~\ref{sec:discussion}. We expect to add them back when such assessments are available in the future.

First, we decided to move the physical actuation factor to the secondary layer because this information is not usually directly related to the privacy and security of the device. In the consumer study, all participants found this information useful from a safety point of view, but none wanted us to move this factor to the primary layer as they reported that the information conveyed by this factor does not have privacy and security implications for them.

Second, we decided to move frequency of data sharing to the secondary layer because prior work has shown that most people do not understand the privacy and security implications of the frequency of data sharing~\cite{balebako2013little}. While almost all the consumers we interviewed were concerned about data sharing, only four mentioned privacy concerns related to the frequency of sharing. 

Third, we decided to exclude device warranty period from the label because it has few, if any, privacy, security, or safety implications.

\vspace{2mm} 

\subsubsection{Secondary Layer}
We found 13 factors that at least 4 out of 7 experts wanted to include on the secondary layer of the label:

\begin{itemize}[leftmargin=*]
    \item {Retention time}
    \item {Purpose of data collection}
    \item {What information can be inferred from the collected data}
    \item {Supported standards (e.g., Wi-Fi, Zigbee)}
    \item {Where the collected data is stored}
    \item {Whether or not the collected data will be linked with data obtained from other sources}
    \item {Special data handling practices for children’s data}
    \item {The control that users are offered (e.g., opt-in/out from data sharing)}
    \item {Data-collection frequency (e.g., once a month, on install)}
    \item {Whether or not the device can still function when Internet connectivity is turned off}
    \item {Relevant security and privacy laws and standards to which the device complies (e.g., ISO 27001, GDPR)}
    \item {Link to the device's key management protocol}
    \item {Resource usage in terms of power and data (e.g., kw, kbps)}
\end{itemize}

Experts mentioned two common reasons to include a factor in the secondary layer rather than the primary layer: the factor requires detailed information to convey risk to consumers (mentioned by 6/7) or the factor does not convey critical information related to the privacy and security of the device (mentioned by 4/7). 

Among the factors our experts wanted to include on the secondary layer, there were three factors that we decided to include on the primary layer instead: date of the latest firmware update, purpose of data collection, and where the collected data is stored.

Experts wanted to have the date of firmware update on the secondary layer mainly because these updates happen frequently and the information on the label can become outdated. On the other hand, consumers need to know the firmware version to which the label is applicable. Therefore, we believe the firmware version number and date information should be provided on both layers of the label.

Most experts (6/7) believed that it would be hard to fit all the purposes of data collection on the primary layer of the label. Therefore, they recommended including this information on the secondary layer. However, past research has shown that purpose of data collection is one of the most important factors consumers want to consider when making privacy decisions~\cite{lee2016understanding, leon2013matters, bonne2017exploring}. Purposes may be grouped into high-level categories that could be included on the primary layer. For example, the W3C's Platform for Privacy Preferences (P3P) standard identified 12 purpose categories~\cite{P3P}. The consumers we interviewed indicated that it was important to them to know the type of data collected and the purpose of collection when making device purchase decisions.  

Experts stated that where the data is stored should be included on the secondary layer because it is not relevant to privacy or security. However, we believe local storage versus data being stored on the cloud can indeed have different privacy and security implications~\cite{sen2015security}. Therefore, we decided to include this in the primary layer. Moreover, most consumer participants were able to reason about privacy and security implications of cloud versus device and discuss the trade-off between security and convenience. C10 talked about this trade-off by saying: 
\begin{quote}
The advantage of the cloud is that if the device is damaged, you can still access it. So it's going to be always available as long as you can access internet. The other issue with the cloud though is that, like, it can be hacked and also, who has access to that is less clear, or you have less control over that. But I can always access it from whatever device I have and it's convenient.
\end{quote}

Although experts recommended including information about device resource usage on the secondary layer, we decided to remove this factor from the label due to its lack of privacy, security, or safety implications. 

\subsubsection{Factors with no Specific Layer}
There were four remaining factors that at least 4 out of 7 experts were enthusiastic about having on the label, but their opinions were split between including them on either the primary or the secondary layer. These factors were: 

\begin{itemize}[leftmargin=*]
    \item {Who the data is shared with}
    \item {Who the data is sold to}
    \item {Whether or not the device can still function when data-driven smart features (e.g., the learning function of smart thermostat) are turned off} 
    \item {Whether or not the device has parental control mode}
\end{itemize}

For these factors, about half of the experts reported that they would like to include them on the primary layer since they are important privacy and security factors that consumers should know about before making purchase decisions. The other half of the experts were not enthusiastic about including these factors on the primary layer. 

We decided to put the factors related to parental controls and device functionality when smart features are turned off on the secondary layer because they are not directly related to security or privacy.

Some experts (3/7) noted that who data is shared with or sold to is likely to change over time, and recommended putting these factors on the secondary layer where they could be updated more easily. We showed consumers a label with these factors on the secondary layer. However, all consumer participants expressed concern when they saw that their information could be shared and sold with third parties and 8 out of 15 said who data is sold to or shared with were among the most important factors that could inform their purchase decisions. Therefore we moved these factors back to the primary layer. 

\subsubsection{Factors to Exclude from the Label}
There were six factors that at least 6 out of 7 experts believed should not be included on the label:

\begin{itemize}[leftmargin=*]
\item {List of device-compatible products}
\item {Link to the device's software and hardware bill of material}
\item {Link to the device's accompanying app(s)}
\item {Whether or not the device manufacturer has a bug bounty program}
\item {Where and when the device brand was incorporated}
\item {Consumer Reports rating}
\end{itemize}

The most common reasons experts said these factors were not suitable for the label were the lack of relevance to privacy and security and inability to convey risk to consumers. For example, S2 did not want the label to include whether or not the manufacturer has a bug bounty program as this factor does not offer adequate insight into security practices of the company: \say{This information is not too important on how the company does security analysis.}

Almost all experts were opposed to including the Consumer Reports rating, mainly as they believed this organization's reputation does not stem from their privacy and security assessments. However, Consumer Reports is in the process of developing a digital privacy and security standard~\cite{digital}, so this may change in the future.

We decided to include whether or not the device manufacturer has a bug bounty program on the secondary layer. Since the word \say{bug bounty} was not immediately clear to consumer participants, we changed the wording to vulnerability disclosure and management, which was more understandable to them. When we presented this factor to our participants, 13 out of 15 associated this information with having good privacy and security practices, hence they were more inclined to trust the company who were transparent about their devices' discovered vulnerabilities and had taken steps to manage them. C3 explained: \say{This factor shows that this is a company that has a security process, and participates in public activities to educate the community on things that can go wrong.} C5 wanted the IoT companies to disclose their devices' history of known vulnerabilities:  \say{A lot of times, if the company had some kind of vulnerability, they maybe want to sweep it under the rug and not let anyone know about it. That's good that it shows you that they're being honest about what issues they've had in the past, and what they've done to address them.} 

We decided to include a link to the software and hardware bill of materials (mentioned on the label as software and hardware composition list) on the secondary layer since it can provide useful information related to security when it is available. When we asked consumers about this factor, most wanted it to be included and noted that even if they did not understand it, it could be useful to those with technical expertise.

Based on our experts' opinions, we initially excluded the list of device-compatible products from the label. In the consumer interview study, we asked participants to tell us about anything they thought was missing from the label that they would like us to add. The only factor that participants suggested was a link to the privacy statement of device-compatible products such as Alexa. As this was suggested by a couple of our early participants, we added a factor on the secondary layer to list compatible platforms with a link to their privacy policies.

\subsection{Attitudes toward Labels and Layered Design}
All consumer participants discussed how difficult it is for them currently to find information related to privacy and security of smart devices before purchasing them. They all reported that they would like to have an IoT security and privacy label available at the point of sale, mainly to be as informed as possible.

Most experts mentioned that IoT privacy and security labels are useful to inform consumers when making purchase decisions, which is in line with prior work~\cite{emami2019exploring}. P7 explained that a label can provide consumers with information they would not have otherwise: 

\begin{quote}
What's good about a label is that it empowers the consumer to make a more active decision about cybersecurity rather than just being completely helpless as to what the security of her device might be. Especially as more and more of this technology is designed for consumers, the average consumer doesn't have a privacy, security, or a legal department to review this stuff before they buy it. Enterprises do, but consumers do not, so someone's gotta be looking out for consumers and giving the consumers this information.
\end{quote}

All consumer participants were familiar with layered labels, as they had seen QR codes on products such as food, drugs, or video games. Most of our participants (11/15) expressed positive attitudes toward the layered design, mainly because the amount of information that could fit in two layers would not fit on a single-layer label. Participants also appreciated the ability to easily gain further insights about the privacy and security practices of the device manufacturer. These participants reported that they engage in a combination of online and in-store shopping. Hence, they believed the layered label design would be useful to them throughout their purchase process.

Some consumer participants (4/15) thought a layered label would not be ideal, citing the inconvenience of using a phone to scan the QR code when shopping in a store, especially for the elderly. C15 explained: \say{These technologies for older generation, they are kind of tough. The idea of installing something to scan the QR code, it's going to be too much for them. I know they prefer to just read everything, put on their glasses and read everything line by line.} Two participants expressed concern that companies might withhold important information from the primary layer and put it only in the secondary layer.  

For each of the factors on the label, we asked consumer participants to tell us how they believe that factor would impact their purchase decisions and whether they would like us to remove the factor or add additional details to the factor. All participants understood the factors presented on the primary layer and were able to discuss the potential risks associated with all of these factors except the level of detail for data storage. Although they all understood the terms ``identifiable'' and ``anonymous,'' participants did not associate identifiable data with risk in this context, perhaps because the utility of a security camera is increased if it can record videos in which people are identifiable. Further testing is needed to understand the impact of the interaction between the purpose of data collection and the granularity of data on consumers' privacy concerns.  

As we expected, some of the information on the secondary layer did not convey privacy and security risks to consumer participants. However, participants still asked to see most of the factors that we included on the secondary layer because they wanted to be as informed as possible when purchasing a smart device. 

Participants mentioned that they might search online for information about unfamiliar factors and the availability of our label would help them. C5 explained: \say{I don't know what TCP and UDP are. But it's interesting to have this here, because then I could go to Reddit and ask on there what that means and what the capabilities are.} 

There were only six factors that 2 or 3 consumer participants thought should be removed from the label. These secondary-layer factors were perceived by those participants as lacking relevance to privacy and security (physical actuations, hardware and software bill of material), not understandable (MUD compliant, key management protocols, open network ports), and not relevant to them (special data handling practices for children). 

At least one consumer participant recognized that some of the factors in the secondary label might be useful to experts. C3 explained: 
\begin{quote}
Labels are both for customers and for experts such as tech journalists, consumer advocacy groups, who are capable of understanding it and who will click on the things, and if they see something that is questionable will raise it in the public press, will raise it with regulatory authorities, and otherwise. The label is not just for the consumer, but also there's another feedback process that works through experts to the extent that the information is available at all.
\end{quote}

\subsection{Prototype Privacy and Security Label}
\begin{figure}[t] 
\includegraphics[trim={0.0in 0.0in 0.0in 0.0in}, clip, width=\columnwidth]{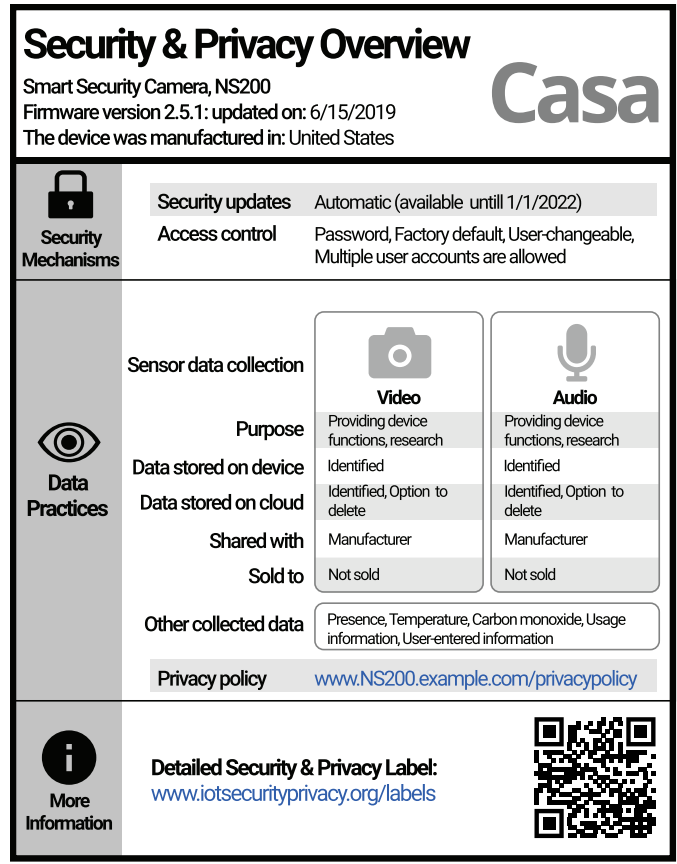}
    \caption{Primary layer of the label. This layer is designed to be printed on product packaging or to appear on a product website. View our latest label design at \protect\url{www.iotsecurityprivacy.org}.}
    \label{fig:primary}
    \vspace{-1em}
\end{figure}

\begin{figure}[t]
\includegraphics[trim={0.0in 0.0in 0.0in 0.0in},clip,width=\columnwidth]{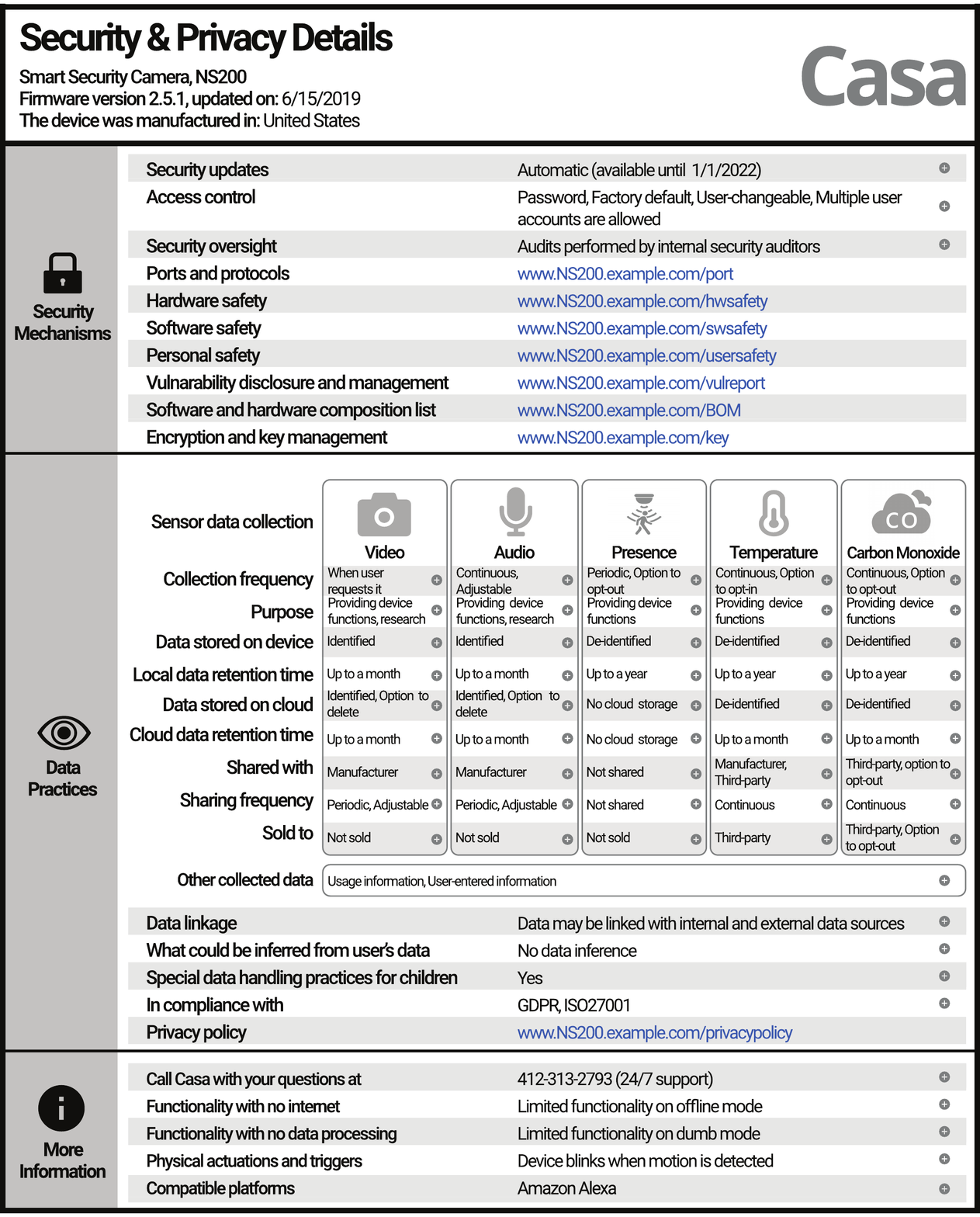}
    \caption{Secondary layer of the label, which can be accessed by scanning the QR code or typing the URL on the primary layer. View our latest label design at \protect\url{www.iotsecurityprivacy.org}.}
    \label{fig:secondary}
    \vspace{-1em}
\end{figure}

We used  the results from our expert elicitation and consumer studies as well as recent IoT security standardization and certification efforts to inform the design of a prototype privacy and security label for a hypothetical smart security camera. Our design has primary and secondary layers, as shown in Figures~\ref{fig:primary} and~\ref{fig:secondary}, respectively. The secondary layer includes plus signs next to each item that can be clicked to reveal further details. We envision that the secondary layer would be accompanied by a computer-readable version of the label to enable automated processing and comparison between products, for example by personal privacy assistants~\cite{colnago2020PPA} or search engines~\cite{egelman2009timing, cranor2016large}. Our website at \url{www.iotsecurityprivacy.org} has the latest label design. 

\section{Discussion} \label{sec:discussion} 
Expert participants recommended including privacy and security star ratings on an IoT label, mostly because ratings would help consumers to more easily compare IoT devices based on their privacy and security practices. All of our consumer participants liked the idea of having privacy and security assessments from trustworthy organizations. Although we believe these third-party assessments would inform consumers' purchase behavior, we decided not to include them on our proposed label, as there is no organization currently doing these evaluations at scale for a wide range of IoT devices. We expect to add a place for assessment information once it is available.

We begin this section by providing a comparison between certifications and star ratings. Next, we discuss possible approaches to IoT privacy and security certifications. 

\subsection{Star Ratings vs. Certification Levels}
Similar to the Energy Star rating system managed by the U.S. Environmental Protection Agency (EPA) and the U.S. Department of Energy (DOE)~\cite{rating4}, the idea of star ratings has been proposed for IoT devices to help consumers make informed purchase decisions~\cite{rating1, rating2}. In a hearing of the U.S. Senate Committee on Commerce, Science, and Transportation’s Subcommittee on Security~\cite{rating3}, Senator Ed Markey suggested a 5-star security rating system for IoT products. In our study, while experts were supportive of privacy and security ratings on the label, they also mentioned two potential challenges of including them.

The first challenge experts brought up relates to the rating scale. Experts suggested that consumers might have trouble distinguishing a large number of ratings, yet a more granular scale could help manufacturers better differentiate their privacy and security practices. P1, who works in industry, discussed this issue: \say{I'm sure industry people, manufacturers, will want more in there. What would happen if you had something like this is it might start to grow based on features they want reflected in that rating. Then I can see it becoming a bigger and bigger scale.} 

Experts mentioned that ratings might pose an unhealthy incentive for IoT companies to achieve full-star ratings only to be able to compete in the market. Companies may be able to game the ratings in order to get all the stars and eventually all products will have all stars, whether they deserve them or not. S2, an academic, explained: \say{The problem I have with ratings like this is that everybody's gonna get a five star, because everybody's gonna figure out how to get the five star.} 

To address these challenges, some experts discussed the idea of having multiple certification levels (e.g., silver, gold, platinum) with a secure baseline or minimum standard instead of star ratings. This is similar to what the LEED standards use for rating energy efficiency and sustainability of buildings~\cite{sustain}. P8 explained: \say{I think consumers should know it passes the minimum security level. If I'm buying a space heater, I know they're not allowed to sell me one that will set on fire. I don't have to say, oh, it has a 70\% score that it will set the house on fire.}

Underwriters Laboratories (UL) published a 5-level IoT security standard (bronze, silver, gold, platinum, and diamond) in 2019~\cite{ul_standard}. As of January 2020, no devices have been certified~\cite{ul_certified}. As manufacturers start having their devices certified, this certification could be added to the IoT label.  

Since the lowest certification level indicates a safe device, there is a risk that manufacturers will aim to achieve the lowest level and not bother pursuing higher levels. Market competition may encourage manufacturers to pursue higher certification levels, especially for devices where the consequences of security breaches are most severe. 

\subsection{Privacy and Security Evaluation and Scoring}
Over the past few years, a number of organizations and research teams have started to develop standards for IoT privacy and security evaluation and scoring. They include Consumer Reports~\cite{digital}, YourThings~\cite{alrawi2019sok, score}, and UL~\cite{ul_standard}.

\subsubsection{Digital Standard}
In 2017, Consumer Reports launched the Digital Standard to work toward providing a comprehensive standard that enables organizations to evaluate consumer IoT products. This standard focuses on four categories: security, privacy, ownership, and governance \& compliance~\cite{digital}. 

The security category of the Digital Standard includes build quality, data security, and personal safety. 

Build quality refers to product stability and whether ``software was built and developed according to the industry's best practices for security.'' The Cyber Independent Testing Lab (CITL)~\cite{cyberitl}, a Digital Standard partner, is actively evaluating and scoring software of IoT devices according to a number of factors. Our label design includes a software safety features element where manufacturers can provide a URL with information related to software security.

Data security includes authentication, encryption, updatability, security audits, and vulnerability disclosure program. All of these factors are included on our label. 

The personal safety category has not yet been defined in the Digital Standard, although developer notes indicate it will be related to avoiding abuse and harassment. Media reports suggest there are many incidents involving smart home devices being used for domestic  abuse~\cite{NYTimes}. However, device manufacturers appear to be doing little to address the risks associated with abuse involving their devices~\cite{ucl}. We have included a factor called personal safety, which provides a place where device manufacturers can indicate available safeguards against abusive behavior once such safeguards have been implemented. Further discussions with experts are needed to determine how to address significant safety issues effectively on the label. As it was explained by S4: \say{Safety means if your car gets hacked, you die. The room that has a laser attached and if it gets hacked, it kills you. A drone can be reprogrammed to dive-bomb your child. I'm not sure how to capture that on the label.} 

The privacy section in the Digital Standard includes user controls, data use and sharing, data retention, and overreach. The assessment procedure for almost all the privacy factors in the Standard involves verifying the company's claimed data practices with actual data practices.

All the privacy factors mentioned in the Digital Standard are covered in our proposed label, except overreach. Overreach, or ``collecting too much data'' focuses on determining whether data collection is beneficial to the user, fully disclosed, the minimum necessary for functionality, and private by default. This seems like an area where a third-party assessment rather than a self report is likely warranted. 

As some of the experts we interviewed mentioned, consumers may weigh privacy and functionality trade-offs differently. Thus it may be difficult to capture a single privacy rating that makes sense for all consumers. In addition to providing detailed information about data practices, a future privacy rating system could be customized based on a consumer's stated privacy preferences, which could change over time.

\subsubsection{YourThings}
Alrawi et al.~\cite{alrawi2019sok} developed a security evaluation and scoring method for smart home devices. In their YourThings~\cite{score} initiative, they produce device scorecards with grades in four areas: device, mobile application, cloud endpoints, and network communication. While our label provides information related to all of the major areas of the YourThings rubric as well as some  security and privacy factors not addressed by YourThings, the YourThings scorecard considers some additional security details, including some that do not lend themselves to self report. The YourThings scorecard offers a concise expert summary of device security issues, which could be useful to include on the label. 

The YourThings rubric considers five device-related factors: upgradability, exposed services, vulnerabilities, configuration, and Internet pairing. We include all on our label. 

The mobile section of the YourThings rubric includes sensitive data, programming issues, and ``over-privileged,'' i.e., requesting excess permissions that are not used or required. Sensitive data is defined in the rubric to include \say{artifacts like API keys, passwords, and cryptographic keys that are hard-coded into the application.} Our label includes factors related to sensitive data and programming issues such as software safety features and key management protocol. However, over-privileged is a factor better assessed by a third-party evaluator rather than being self-reported.

The cloud endpoints section of the rubric includes domain categories, TLS configuration, and vulnerable services. Some of the information needed to compute this score is included on our secondary layer of the label when fully expanded. However, details needed to evaluate TLS configuration are not included.

Finally, the network communication section of the rubric includes protocols, susceptibility to Man in the Middle (MITM) attack, and use of encryption. While the secondary layer of our label provides some of the information needed to compute this score, a third-party evaluation would be needed to provide a complete assessment.

The concise YourThings scores are useful for comparing devices, but users may need to drill down to obtain information relevant to their specific needs. For example, devices are penalized for not having automatic updates. While automatic updates are generally considered the most secure approach, poorly timed updates can be problematic, potentially interfering with critical device functions. 

\subsubsection{UL}
The 5-level UL certification process includes 44 requirements over seven categories: software updates, data \& cryptography, logical security, system management, customer identifiable data, protocol security, and process \& documentation~\cite{ul_method}.

While our proposed label includes factors from all seven categories, a third-party evaluation is needed to assess compliance with requirements. Our label can inform consumers about security and privacy, and goes into more detail about privacy issues than UL's customer identifiable data category. By including the UL certification in our label, we would offer users a single concise assessment of device security that complements the more detailed information provided on the label.

\section{Conclusion}
We conducted a study with 22 privacy and security experts to elicit their opinions on the contents of IoT privacy and security labels. By following a three-round Delphi method, we found the factors that experts believed should be included on the label, and distributed them between primary and secondary layers of the label in three categories (security mechanisms, data practices, and general/more information). By conducting a series of in-depth semi-structured interviews with 15 IoT consumers, we iteratively improved the design of our proposed privacy and security label for IoT devices. Additional user tests should explore how consumers use the label in context as well as identify areas where wording or design can be further improved. Finally, we plan to develop a glossary of terms for consumers and an implementation guide for device manufacturers with detailed definitions of each factor and its possible values. The latest version of the label and implementation information, as well as a tool for generating the label are available at \url{www.iotsecurityprivacy.org}.

\section*{Acknowledgments}
We thank Supawat Vitoorapakorn and Shreyas Nagare for their help in designing the labels and the project website. This research has been supported in part by DARPA and the Air Force Research Laboratory FA8750-15-2-0277, and NSF awards TWC-1564009 and SaTC-1801472. The US Government is authorized to reproduce and distribute reprints for Governmental purposes not withstanding any copyright notation thereon. Additional support has also been provided by Google and by the Carnegie Mellon CyLab Security and Privacy Institute. The views and conclusions contained herein are those of the authors and should not be interpreted as necessarily representing the official policies or endorsements, either expressed or implied, of DARPA, the Air Force Research Laboratory, the US Government, or Google.

\interlinepenalty=10000

\balance

\Urlmuskip=0mu plus 1mu\relax
\def\UrlBreaks{\do\/\do_\do-}

\bibliographystyle{abbrv}
\bibliography{bibliography}

\appendices
\twocolumn[\section{Expert Elicitation Study}]\label{sec:expert_appendix}

\renewcommand{\thefigure}{7}
\begin{figure}[h]
\vspace{3em}
\tcbox[boxsep=0mm]{\includegraphics[trim={0in 4.02in 0in 0in}, clip, width=.9\columnwidth]{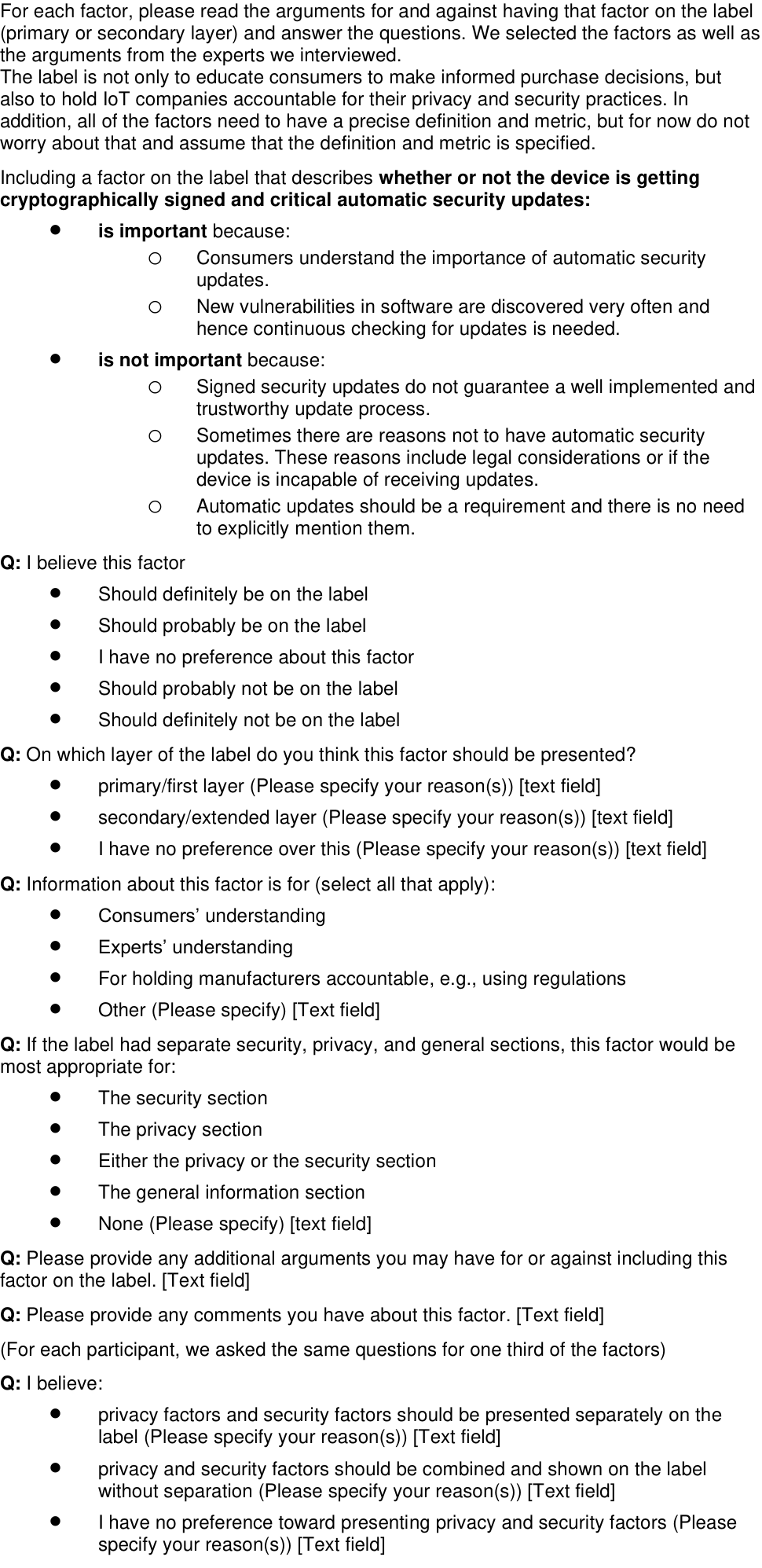}}
    \caption{Questions we asked on the second survey. Here we presented experts with common arguments from the first survey and asked them again to make decision about each and provide us with their additional arguments.}
    \label{fig:survey2}
\end{figure}

\renewcommand{\thefigure}{5}
\begin{figure}[!h]
\vspace{3em}
\tcbox[boxsep=0mm]{\includegraphics[trim={0in 8.3in 0in 0in}, clip, width=.9\columnwidth]{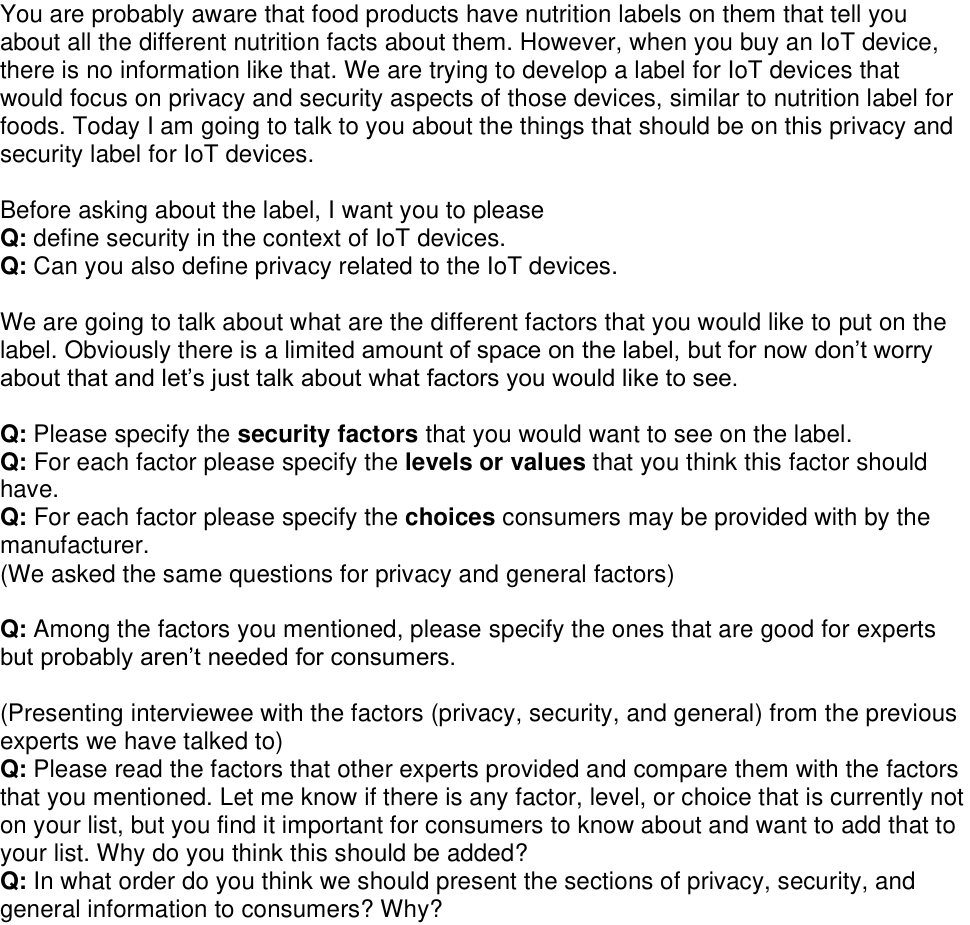}}
    \caption{Questions we asked in our semi-structured interviews with 22 experts. Based on the interviews, we found 47 factors that experts expressed interest in including on IoT privacy and security label.}
    \label{fig:interview}
    \vspace{.78em}
\end{figure}

\renewcommand{\thefigure}{6}
\begin{figure}[!h]
\tcbox[boxsep=0mm]{\includegraphics[trim={0in 9in 0in 0in}, clip, width=.9\columnwidth]{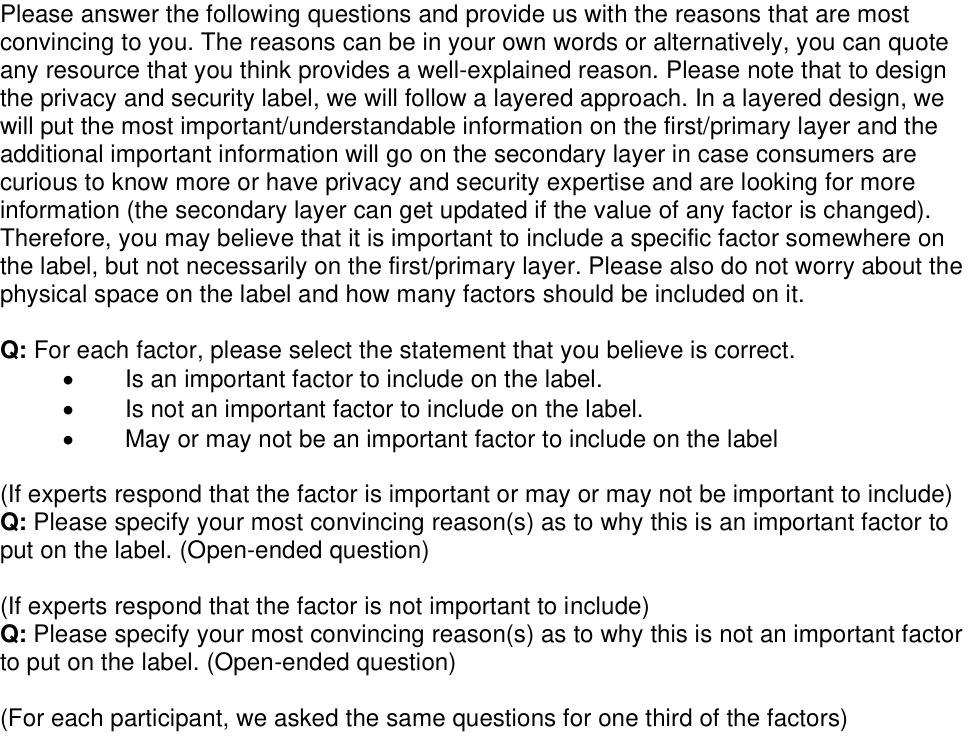}}
    \caption{Questions we asked on the first survey. In this stage, we asked experts to review the factors from the interview study and specify their preferences about including each factor, as well as their reasons.}
    \label{fig:survey1}
    \vspace{-7em}
\end{figure}

\twocolumn[\section{Consumer Study}]\label{sec:consumer_appendix}

\begin{minipage}{\textwidth}
\vspace{.5in}
\renewcommand{\thefigure}{8}
\tcbox[boxsep=0mm]{\includegraphics[trim={2.4in 5.25in 2.4in 0in}, clip, width=.45\columnwidth]{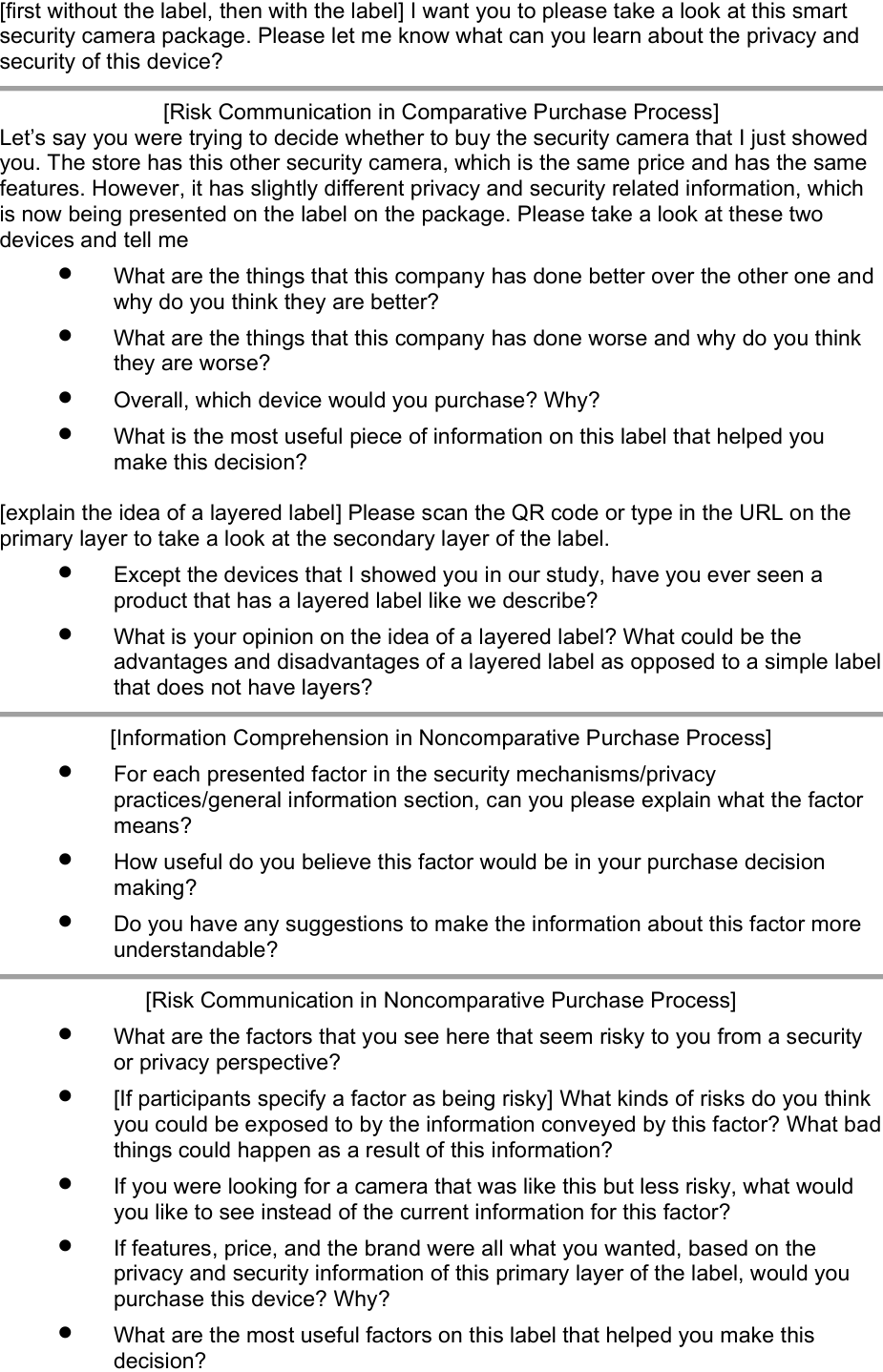} \includegraphics[trim={2.4in -.56in 2.4in 5.75in}, clip, width=.45\columnwidth]{drawing-1.pdf}}
\captionof{figure}{After the expert elicitation study and specifying the content of the IoT privacy and security label based on experts' input, we conducted in-depth semi-structured interviews with 15 participants, who had purchased at least one smart home device. We asked interviewees questions to study how much they understand the content of our designed labels and how the presented information conveyed risk to participants.}
\label{fig:consumer_interview}
\end{minipage}

\end{document}